\documentclass[prl,twocolumn,preprintnumbers,nofootinbib]{revtex4}
\usepackage[a4paper, hdivide={1.91cm,,1.165cm}, vdivide={1.83cm,,3.6cm}]{geometry}

\usepackage{amsmath,amssymb}
\usepackage{graphicx,multirow}
\usepackage{color}
\usepackage{units}
\usepackage[hyperfootnotes=false]{hyperref}

\def \L {\mathcal{L}} 

\def \vec#1{{\boldsymbol{#1}}}
\newcommand{\hc}{\ensuremath{\text{h.c.}}}

\allowdisplaybreaks

\begin{document}

\title{Interpretation of Lepton Flavor Violation}

\preprint{ULB-TH/16-19}

\author{Julian \surname{Heeck}}
\email{Julian.Heeck@ulb.ac.be}
\affiliation{Service de Physique Th\'eorique, Universit\'e Libre de Bruxelles, Boulevard du Triomphe, CP225, 1050 Brussels, Belgium}

\hypersetup{
    pdftitle={Interpretation of Lepton Flavor Violation},
    pdfauthor={Julian Heeck}
}


\begin{abstract}
The observation of a charged-lepton flavor violating process would be a definite sign for physics beyond the Standard Model, but would actually only prove that one particular linear combination of lepton numbers is violated. We categorize lepton-flavor-violating processes by their quantum numbers and show how their discovery can be interpreted model independently, studying in particular which processes are required to establish that the entire flavor group is broken. We also comment on total lepton number, seeing as lepton number violation practically implies lepton flavor violation as well.
\end{abstract}

\maketitle


\section{Introduction}

The classical Lagrangian of the Standard Model (SM) of particle physics features the global symmetry group $U(1)_B\times U(1)_{L_e}\times U(1)_{L_\mu}\times U(1)_{L_\tau}$ associated with the conservation of baryon number $B$ and the three lepton numbers $L_{e,\mu,\tau}$. Defining the total lepton number $L\equiv L_e+L_\mu+L_\tau$, this global symmetry group can also be written in the following basis,
\begin{align}
U(1)_{B+L}\times U(1)_{B-L} \times U(1)_{L_\mu-L_\tau}\times U(1)_{L_\mu+ L_\tau -2 L_e}\,.
\label{eq:global_symmetry}
\end{align}
In principle this is useful because the linear combination $B+L$ is actually broken at the quantum level by non-perturbative processes by six units, specifically $\tfrac13\Delta B=\Delta L_e = \Delta L_\mu = \Delta L_\tau=1$~\cite{'tHooft:1976up}.
Neutrino oscillations have proven furthermore that lepton flavor, $U(1)_{L_\mu-L_\tau}\times U(1)_{L_\mu+ L_\tau -2 L_e}$, is \emph{not} conserved in nature, leaving at most $U(1)_{B-L}$ as an unbroken symmetry~\cite{Heeck:2014zfa}.

Practically speaking, however, the $B+L$ violating processes are much too suppressed (at zero temperature) to ever be observable, and the same goes for \emph{charged}-lepton flavor violation (CLFV) induced by non-zero neutrino masses $m_j$ and a non-trivial leptonic mixing matrix~$U$. For example, Dirac neutrinos lead to $\Delta (L_\beta-L_\alpha)=2$ CLFV at the one-loop level~\cite{Petcov:1976ff},
\begin{align}
\frac{\Gamma (\ell_\alpha\to \ell_\beta \gamma)}{\Gamma (\ell_\alpha \to \ell_\beta \nu_\alpha \bar\nu_\beta)} \simeq \frac{3\alpha_\text{EM}}{32\pi} \left|\sum_{j=2,3} U_{\alpha j} \frac{ \Delta m_{j1}^2}{M_W^2} U_{j\beta}^\dagger\right|^2,
\end{align}
which is smaller than $10^{-53}$ for all channels and hence completely unobservable. The Glashow--Iliopoulos--Maiani mechanism ensures that \emph{all} neutrino-induced CLFV processes are suppressed by the sub-eV$^2$ neutrino-mass-squared differences $\Delta m_{ij}^2$, because degenerate (Dirac) neutrino masses would render $U$ unphysical and thus reinstate lepton flavor symmetry.

As a result, the group of Eq.~\eqref{eq:global_symmetry} is still an excellent \emph{approximate} symmetry for charged leptons; the observation of CLFV would imply physics beyond the SM -- and beyond neutrino oscillations -- with many models being able to saturate current limits (see e.g.~Refs.~\cite{deGouvea:2013zba,Bernstein:2013hba} for current reviews). Processes under experimental investigation are listed in Tabs.~\ref{tab:lfv}--\ref{tab:lfv2}, focusing on rare decays rather than collider signatures.
The next decade will see significant improvement of these limits or even a discovery, with MEG-II, Mu3e, Mu2e, COMET and DeeMe probing $\mu\to e\gamma, 3e$ and $\mu\to e$ conversion; LHCb, BES-III and Belle-II probing CLFV decays of taus and hadrons; and the (HL-)LHC probing CLFV decays of the Brout--Englert--Higgs boson $h$ among other channels. (Limits on LFV $Z$ decays could be improved by many orders of magnitude at future $e^+e^-$ colliders, not listed in our tables.)
It is thus timely to study how a possible discovery can be interpreted.\footnote{At present there is one tantalizing $\sim 2.5\sigma$ hint for CLFV in the channel $h\to\bar\mu\tau+\bar\tau\mu$ by CMS~\cite{Khachatryan:2015kon}, yet to be confirmed or excluded by $\sqrt{s}=\unit[13]{TeV}$ data~\cite{CMS:2016qvi} or ATLAS~\cite{Aad:2015gha,Aad:2016blu}.}
In the following, we will make an effort to study CLFV model independently by focusing on the quantum numbers of the various processes.

\section{Lepton Flavor Violation}

Following standard convention, we define CLFV as processes that conserve total lepton number $L$ (and $B$) but violate $U(1)_{L_\mu-L_\tau}\times U(1)_{L_\mu+ L_\tau -2 L_e}$ and do not involve neutrinos.  The decays in Tabs.~\ref{tab:lfv}--\ref{tab:lfv2} have been sorted into (six) groups according to the lepton numbers that are violated; this already enables us to make model-\emph{independent} qualitative predictions: if \emph{one} process of a given group is observed, \emph{all} processes of said group unavoidably exist, being at least generated at loop level.\footnote{This statement has to be modified if \emph{light} new-physics modes such as $\ell_\alpha\to\ell_\beta Z'$~\cite{Foot:1994vd,Heeck:2016xkh,Altmannshofer:2016brv} are included in the list (or observed).} All processes of one group provide the same quantum-number information. The different branching ratios within each group are model \emph{dependent}, as is the question of whether more than one group is observable.
(The chiralities of the fermions involved in CLFV~\cite{Davidson:2011aa} and the impact on lepton-flavor conserving observables such as $g-2$~\cite{Lindner:2016bgg} are also model dependent.)

\begin{ruledtabular}
\begin{table}
	\centering
		\begin{tabular}{clrr}
		Group & Process & Current  & Future\\		
		\hline
		\multirow{7}{*}{\rotatebox[origin=c]{90}{$\Delta (L_e -L_\mu) =2$}}	& $\mu\to e\gamma$ & $4.2\times 10^{-13}$~\cite{TheMEG:2016wtm} & $4\times 10^{-14}$\cite{Baldini:2013ke}\\
															& $\mu\to e\bar e e$ & $1.0\times 10^{-12}$~\cite{Bellgardt:1987du} & $10^{-16}$~\cite{Blondel:2013ia}\\
															& $\mu\to e$ conv. & $\mathcal{O}( 10^{-12})$~\cite{Bertl:2006up} & $10^{-17}$~\cite{Kuno:2013mha,Bartoszek:2014mya}\\
															& $h\to e \bar\mu$ & $3.5\times 10^{-4}$~\cite{Khachatryan:2016rke} & $2\times 10^{-4}$~\cite{Banerjee:2016foh}\\
															& $Z\to e \bar\mu$ & $7.5\times 10^{-7}$~\cite{Aad:2014bca} & --\\
															& had$\to e \bar \mu$(had) & $4.7\times 10^{-12}$~\cite{Ambrose:1998us} & $10^{-12}$~\cite{Moulson:2013oga} \\
	\hline
		\multirow{8}{*}{\rotatebox[origin=c]{90}{$\Delta (L_e -L_\tau) = 2$}}	 & $\tau\to e\gamma$ & $3.3\times 10^{-8}$~\cite{Aubert:2009ag} & $10^{-9}$~\cite{Aushev:2010bq}\\
															& $\tau\to e\bar e e$ & $2.7\times 10^{-8}$~\cite{Hayasaka:2010np} & $10^{-9}$~\cite{Aushev:2010bq}\\
															& $\tau\to e\bar \mu \mu$ & $2.7\times 10^{-8}$~\cite{Hayasaka:2010np} & $10^{-9}$~\cite{Aushev:2010bq}\\
															& $\tau\to e\,$had & $\mathcal{O}(10^{-8})$~\cite{Olive:2016xmw} & $10^{-9}$~\cite{Aushev:2010bq}\\
															& $h\to e\bar \tau$ & $6.9\times 10^{-3}$~\cite{Khachatryan:2016rke} & $5\times 10^{-3}$~\cite{Banerjee:2016foh}\\
															& $Z\to e\bar \tau$ & $9.8\times 10^{-6}$~\cite{Akers:1995gz} & --\\
															& had$\to e\bar \tau$(had) & $\mathcal{O}(10^{-6})$~\cite{Ablikim:2004nn,Lees:2010jk} & -- \\
	\hline
		\multirow{8}{*}{\rotatebox[origin=c]{90}{$\Delta (L_\mu -L_\tau) = 2$}} & $\tau\to \mu\gamma$ & $4.4\times 10^{-8}$~\cite{Aubert:2009ag} & $10^{-9}$~\cite{Aushev:2010bq}\\
															& $\tau\to \mu\bar e e$ & $1.8\times 10^{-8}$~\cite{Hayasaka:2010np} & $ 10^{-9}$~\cite{Aushev:2010bq}\\
															& $\tau\to \mu\bar \mu \mu$ & $2.1\times 10^{-8}$~\cite{Hayasaka:2010np} & $10^{-9}$~\cite{Aushev:2010bq}\\
															& $\tau\to \mu\,$had & $\mathcal{O}(10^{-8})$~\cite{Olive:2016xmw} & $ 10^{-9}$~\cite{Aushev:2010bq}\\
															& $h\to \mu\bar \tau$ & $1.2\times 10^{-2}$~\cite{CMS:2016qvi} & $5\times 10^{-3}$~\cite{Banerjee:2016foh}\\
															& $Z\to \mu\bar \tau$ & $1.2\times 10^{-5}$~\cite{Abreu:1996mj} & --\\
															& had$\to \mu\bar \tau$(had) & $\mathcal{O}(10^{-6})$~\cite{Ablikim:2004nn,Lees:2010jk} & -- \\
		\end{tabular}
		\caption{CLFV with conserved $L$ and $B$, omitting CP conjugate processes. Current limits on the branching ratios are at $90\%$~C.L.~($h$/$Z$ decays at $95\%$~C.L.). A full list of CLFV involving hadrons (had) can be found in the PDG~\cite{Olive:2016xmw}.
		\label{tab:lfv}}
\end{table}
\end{ruledtabular}

\begin{ruledtabular}
\begin{table}
	\centering
		\begin{tabular}{llrr}
		Group & Process & Current  & Future\\		
		\hline
		$\Delta (L_\mu +L_\tau-2 L_e) = 6$ & $\tau\to ee\bar \mu$ & $1.5\times 10^{-8}$~\cite{Hayasaka:2010np} & $10^{-9}$~\cite{Aushev:2010bq}\\
	\hline
		$\Delta (L_\tau+L_e-2 L_\mu) = 6$ & $\tau\to \mu\mu\bar e$ & $1.7\times 10^{-8}$~\cite{Hayasaka:2010np} & $10^{-9}$~\cite{Aushev:2010bq}\\
	\hline
		$\Delta (L_e+L_\mu -2L_\tau) = 6$ & $\mu e \to \tau\tau$ & -- & --
		\end{tabular}
		\caption{CLFV with conserved $L$ and $B$, omitting CP conjugate processes. Current limits at $90\%$~C.L.
		\label{tab:lfv2}}
\end{table}
\end{ruledtabular}

It is indeed possible that only one of the groups in Tabs.~\ref{tab:lfv}--\ref{tab:lfv2} is observable, i.e.~only one linear combination of lepton flavors is violated, while the others are conserved (outside of the neutrino sector). A concrete model has been put forward in Ref.~\cite{Heeck:2014qea} that only generates the $\Delta (L_\mu-L_\tau)=2$ group of Tab.~\ref{tab:lfv} by extending a two-Higgs-doublet model with a spontaneously broken $U(1)_{L_\mu-L_\tau}$ symmetry. Models for $\Delta (L_e - L_{\mu,\tau})=2$ can be built in complete analogy.
The groups of Tab.~\ref{tab:lfv2}, which consist of at most one observable process, can for example be obtained by extending the SM by an $SU(2)_L$ singlet scalar $k^{++}$ with electric charge 2, which has the Yukawa couplings
\begin{align}
\L \supset g_{\alpha\beta} \overline{\ell}_\alpha^c P_R \ell_\beta\, k^{++} +\hc ,
\label{eq:kplusplus}
\end{align}
with symmetric coupling matrix $g_{\alpha\beta}$ and chiral projection operator $P_R$~\cite{Babu:1988ki}. Imposing a (local or global) $U(1)_{L_\mu - L_\tau}$ symmetry severely restricts $g$,
\begin{align}
\begin{split}
\L &\supset \left(g_{\mu\tau} \overline{\mu}^c_R  \tau_R+ g_{\tau\mu} \overline{\tau}^c_R \mu_R+g_{ee} \overline{e}^c_R e_R\right) k^{++} +\hc ,
\end{split}
\end{align}
thus allowing only for the $\Delta (L_\mu+L_\tau-2 L_e) = 6$ decay $\tau\to ee\bar\mu$ of Tab.~\ref{tab:lfv2} but none of the other CLFV groups. 
(Note that $L$ is conserved if we assign $L(k^{++}) = -2$, so the above Lagrangian by itself does not lead to (Majorana) neutrino mass.)
The two other $\Delta (L_\alpha + L_\beta - 2 L_\gamma)=6$ groups of Tab.~\ref{tab:lfv2} can be generated analogously by imposing $U(1)_{L_\alpha - L_\beta}$ on Eq.~\eqref{eq:kplusplus}.
It is not difficult to construct other models that generate only \emph{one} of the groups of Tabs.~\ref{tab:lfv}--\ref{tab:lfv2} by making use of the symmetries involved, but this should suffice as a proof of principle.

The above discussion is meant to illustrate the somewhat trivial point that the observation of \emph{one} CLFV process only proves that \emph{one} linear combination of lepton numbers is broken, while the other(s) might still be conserved. Using without loss of generality the basis of Eq.~\eqref{eq:global_symmetry}, the observation of one CLFV process means that $U(1)_{L_\mu-L_\tau}\times U(1)_{L_\mu+ L_\tau -2 L_e}$ is broken to a $U(1)'\times \mathbb{Z}_N$ subgroup. The observation of a \emph{second} CLFV process (from a different group) implies that $U(1)'\times \mathbb{Z}_N$ is further broken down to 
\begin{align}
\mathbb{Z}_{\Delta (L_\mu-L_\tau)}\times \mathbb{Z}_{\Delta (L_\mu+ L_\tau -2 L_e)}\,,
\label{eq:Znsubgroup}
\end{align}
which might contain some redundancies as we will see later (see also Ref.~\cite{Petersen:2009ip} on this topic).
All CLFV processes can be conveniently drawn on this lattice, shown in Fig.~\ref{fig:LFV_grid}, labeled by one representative process (e.g.~$\mu\to e\gamma$ stands for all $\Delta (L_e-L_\mu)=2$ processes from Tab.~\ref{tab:lfv}).
Here we included far more processes than listed in Tabs.~\ref{tab:lfv}--\ref{tab:lfv2} for the sake of illustration, even though many of them are not testable.
In fact, the only realistically testable processes not already listed are $\bar\mu e\to \mu\bar e$ (violating $\Delta (L_e-L_\mu)=4$), for which experimental limits from muonium exist~\cite{Willmann:1998gd}, and $\tau\to\mu\mu\mu\bar e\bar e$.

\begin{figure*}[t]
\includegraphics[width=0.75\textwidth]{LFV_grid}
\caption{
CLFV processes (only one representative shown per group) organized by their $U(1)_{L_\mu-L_\tau}\times U(1)_{L_\mu+ L_\tau -2 L_e}$ breaking.
}
\label{fig:LFV_grid}
\end{figure*}

Simple vector addition now allows us to make model-\emph{independent} interpretations of CLFV from Fig.~\ref{fig:LFV_grid}. For example, the observation of $\mu\to e\gamma = (-3,-1)$ and $\tau\to \mu \gamma = (0,2)$ implies that $\tau\to e\gamma = (-3,1)$ must exist, as well as in fact \emph{all} other CLFV processes, because each point on the lattice can be written as $n (-3,-1)+m(0,2)$ with $n,m\in \mathbb{Z}$. Observing two different groups of Tab.~\ref{tab:lfv} is thus sufficient to prove that all CLFV exist, i.e.~that the flavor group $U(1)_{L_\mu-L_\tau}\times U(1)_{L_\mu+ L_\tau -2 L_e}$ is completely broken.\footnote{From Fig.~\ref{fig:LFV_grid} it might seem like $U(1)_{L_\mu+ L_\tau -2 L_e}$ always has an unbroken $\mathbb{Z}_3$ subgroup under which $(\mu,\tau,e)\sim (1,1,-2)$, but this is actually a $U(1)_L$ subgroup because $(1,1,-2) = (1,1,1)\mod 3$, which can hence be ignored.} It is of course impossible to make model-independent statements about the size and ratios of the different CLFV channels.

\begin{table*}
	\centering
	\renewcommand{\arraystretch}{1.43}
		\begin{tabular}{llll}
		\hline\hline
		Observation of charged lepton flavor violation & $\Rightarrow$ & & Remaining symmetry\\		
	\hline
		$\Delta (L_\alpha-L_\beta)=2$ & & &  $U(1)_{L_\alpha + L_\beta - 2 L_\gamma}$\\
		$\Delta (L_\alpha + L_\beta - 2 L_\gamma)=6$ & & &  $U(1)_{L_\alpha-L_\beta}$\\
	  $\Delta (L_\alpha + L_\beta - 2 L_\gamma)=6$ and $\Delta (L_\alpha-L_\beta)=2$ & & &  $\mathbb{Z}_2$: $\ell_\gamma\to-\ell_\gamma$ \\
		$\Delta (L_\alpha + L_\beta - 2 L_\gamma)=6$ and $\Delta (L_\alpha + L_\gamma - 2 L_\beta)=6$ & & & $\mathbb{Z}_3$: $(\ell_\alpha,\ell_\beta,\ell_\gamma)\sim (0,1,2)$\\
		$\Delta (L_\alpha-L_\beta)=2$ and $\Delta (L_\alpha-L_\gamma)=2$ & & & -- \\
		$\Delta (L_\alpha-L_\beta)=2$ and $\Delta (L_\alpha + L_\gamma - 2 L_\beta)=6$ & & & --\\
		\hline\hline
		\end{tabular}
		\caption{Observation of CLFV ($U(1)_{L_\mu-L_\tau}\times U(1)_{L_\mu+ L_\tau -2 L_e}$ breaking) via processes of Tabs.~\ref{tab:lfv}--\ref{tab:lfv2} and remaining subgroups after up to two CLFV discoveries. \emph{Three} observations -- at least one from each table -- imply full flavor breakdown.
		\label{tab:flowchart}}
\end{table*}

Two (orthogonal) CLFV processes are however not \emph{always} enough to establish the full breakdown of the flavor group. For example, the observation of $\tau\to \mu \gamma = (0,2)$ and $\tau\to e e\bar\mu  = (-6,0)$ only implies that a coarser sublattice is generated, which in particular does not include $\mu\to e\gamma$ or $\tau\to e \gamma$ (Fig.~\ref{fig:LFV_grid}). Formally, $U(1)_{L_\mu-L_\tau}\times U(1)_{L_\mu+ L_\tau -2 L_e}$ is broken to a $\mathbb{Z}_2$ subgroup under which electrons flip sign. A \emph{third} CLFV observation, with an odd number of electrons, e.g.~$\mu\to e\gamma$ or $\tau\to \mu\mu\bar e$, is thus necessary to establish full flavor breaking.
For the convenience of the reader, we provide a simple flowchart of possible CLFV in Tab.~\ref{tab:flowchart}, assuming only the processes of Tabs.~\ref{tab:lfv}--\ref{tab:lfv2} are observable. Since we are only able to experimentally test flavor breaking by very few units, the remaining discrete subgroups, i.e.~Eq.~\eqref{eq:Znsubgroup}, are surprisingly simple (only $\mathbb{Z}_2$ or $\mathbb{Z}_3$).
 
Let us make one final remark about the similarity of Fig.~\ref{fig:LFV_grid} with the well-known hadron multiplets of Gell-Mann's \emph{Eightfold Way}~\cite{Gell-Mann:101798}. The latter is a way to organize $q\bar q$ (and $qqq$) states according to their transformation properties under the (approximate) flavor symmetry $SU(3)_f$ with $q = (u,d,s)\sim\vec{3}$, leading in particular to a meson octet as a result of $\vec{8}\subset \vec{3}\otimes\bar{\vec{3}}$. In Fig.~\ref{fig:LFV_grid} we are effectively organizing processes such as $\ell\to \ell(\bar\ell\ell)$ according to their transformation properties under the flavor symmetry $SU(3)_\ell$ with $\ell = (e,\mu,\tau)\sim\vec{3}$. The $\ell_\alpha\to\ell_\beta\gamma$ processes then form a LFV ``octet'' similar to the meson octet, where we of course neglect the two neutral states (corresponding to $\pi^0$ and $\eta$) that do not violate flavor. Similarly, the 12 processes $\ell\to \ell(\ell\bar\ell)$, e.g.~$\tau \to \mu\mu\bar e$ and $e e\to\tau\tau$, can be seen as part of the $\vec{27}\subset \vec{3}\otimes\bar{\vec{3}}\otimes \vec{3}\otimes\bar{\vec{3}}$, which again includes many singlets not of interest for LFV.
Baryon multiplets do not have an analogue in LFV because angular momentum and the assumed baryon number conservation forbid processes such as $\ell\to\bar\ell \bar\ell$ that would correspond to $\vec{3}\otimes \vec{3}\otimes\vec{3}$.
We stress that the similarity of Fig.~\ref{fig:LFV_grid} with the meson multiplets is purely formal; $SU(3)_\ell$ is broken badly by the different lepton masses, leaving only the abelian Cartan subgroup $U(1)_{L_\mu-L_\tau}\times U(1)_{L_\mu+ L_\tau -2 L_e}$ as a possible symmetry of nature, which is thus a better starting point for LFV.

\section{Lepton Number Violation}

For the study of CLFV we assumed $L$ and $B$ to be conserved, which is a convenient simplification -- allowing us to represent CLFV in a two-dimensional plane (Fig.~\ref{fig:LFV_grid}) -- and could well be true if neutrinos are Dirac particles and $B-L$ is unbroken~\cite{Heeck:2014zfa}. Let us loosen this assumption and allow for lepton number violation (LNV), still keeping $B$ conserved for simplicity.
Even though our decomposition of Eq.~\eqref{eq:global_symmetry} suggests LFV (as defined above) and LNV to be unrelated issues, all currently probed LNV processes (Tab.~\ref{tab:lnv}) in fact violate lepton flavor. \emph{True} LNV, conserving flavor and baryon number, is a lot harder to come by and involves at least six leptons -- e.g.~$\bar e\bar e\to \mu\mu\tau\tau$ plus charged bosons -- because all known fermions carry flavor or baryon number.\footnote{Lorentz-invariant operators that conserve $B$ but break $L$ take the form $\ell^{2n}$ times bosonic fields, where $n\in \mathbb{N}$. Since $\ell\sim \vec{3}$ under the $SU(3)_\ell$ flavor group, $n=3$ is the lowest-dimensional operator that contains flavor singlets, seeing as $\vec{3}^6 \supset \vec{1}$.} As such, any observation of LNV is practically also an observation of LFV.
We list possible LNV processes in Tab.~\ref{tab:lnv}, focusing again on decays rather than collider signatures.
We urge experimentalists to complete Tab.~\ref{tab:lnv} by looking for meson decays into $\tau^+ \ell^+$, e.g.~$B^+\to \pi^- \tau^+\ell^+$, with $\ell\in\{e,\mu,\tau\}$~\cite{Zuber:2000vy}.
Similar to CLFV we only list LNV by a few units because experimental signatures of $\Delta L > 2$ are much more challenging~\cite{Heeck:2013rpa,Heeck:2015pia}. (A first preliminary limit on $\Delta L_e = 4$ was presented recently by NEMO-3 in the form of a lower limit of
$\unit[2.6\times 10^{21}]{yr}$ on the neutrinoless quadruple beta decay~\cite{Heeck:2013rpa} ${}^{150}_{60}\mathrm{Nd} \to {}^{150}_{64}\mathrm{Gd}+4 e^-$~\cite{nemo}.)

\begin{ruledtabular}
\begin{table}
	\centering
		\begin{tabular}{llll}
		Group & Process & Current  & Future\\		
	\hline
		$\Delta L_e = 2$ 	& $0\nu \beta\beta$ & $\mathcal{O}(\unit[10^{25}]{yr})$~\cite{Dell'Oro:2016dbc} & $\unit[10^{26}]{yr}$~\cite{Dell'Oro:2016dbc}\\
											& had$\to ee\,$had & $6.4\times 10^{-10}$~\cite{Appel:2000tc} & $10^{-12}$~\cite{Moulson:2013oga}\\
	\hline
		$\Delta L_\mu = 2$ & had$\to \mu\mu\,$had & $8.6\times 10^{-11}$~\cite{Massri:2016dff} & $ 10^{-12}$~\cite{Moulson:2013oga}\\
	\hline
		$\Delta L_\tau = 2$ & had$\to \tau\tau\,$had & -- & --\\
	\hline
		$\Delta (L_e +L_\mu)= 2$ 	& $\mu \to \bar e$ conv. & $3.6\times 10^{-11}$~\cite{Kaulard:1998rb} & $\ll 10^{-11}$~\cite{Geib:2016atx}\\
															& had$\to \mu e\,$had & $5.0\times 10^{-10}$~\cite{Appel:2000tc} & $10^{-12}$~\cite{Moulson:2013oga}\\
	\hline
		$\Delta (L_e+L_\tau) = 2$ & $\tau\to \bar e\,$had & $2.0\times 10^{-8}$~\cite{Miyazaki:2012mx} & $10^{-9}$~\cite{Aushev:2010bq}\\
															& had$\to \tau e\,$had & -- & --\\
	\hline
		$\Delta (L_\mu+L_\tau) = 2$ & $\tau\to \bar \mu\,$had & $3.9\times 10^{-8}$~\cite{Miyazaki:2012mx} & $10^{-9}$~\cite{Aushev:2010bq}\\
															& had$\to \tau\mu\,$had & -- & --
		\end{tabular}
		\caption{Processes violating total lepton number $L$ by two units ($90\%$~C.L.~limits), assuming conserved baryon number.
		\label{tab:lnv}}
\end{table}
\end{ruledtabular}

Of the processes in Tab.~\ref{tab:lnv}, neutrinoless double beta decay ($0\nu\beta\beta$)~\cite{Rodejohann:2011mu,Dell'Oro:2016dbc} is the only one that is sensitive to \emph{neutrino-induced} LNV/LFV. Since we already know from neutrino oscillations that lepton flavor is broken, the observation of $0\nu\beta\beta$ would prove that \emph{all three} lepton numbers $L_{e,\mu,\tau}$ are broken in the neutrino sector. This implies in particular that neutrinos are Majorana particles~\cite{Schechter:1981bd} -- and the existence of additional particles such as scalars or heavier Majorana partners as a UV completion of the Weinberg operator~\cite{Weinberg:1979sa} -- but is not sufficient to tell us anything about \emph{charged} lepton flavor.

An observation of a process from Tab.~\ref{tab:lnv} other than $0\nu\beta\beta$ would on the other hand imply that lepton flavor is definitely broken in the charged-lepton sector,
\begin{align}
U(1)_{L_e} \times U(1)_{L_\mu}\times U(1)_{L_\tau} \to U(1)\times U(1)'\times \mathbb{Z}_N .
\end{align}
For example, the discovery of $\Delta (L_\alpha+L_\beta) = 2$ would imply that $U(1)_{L_\alpha-L_\beta}\times U(1)_{L_\gamma}$ could still be a good symmetry for charged leptons, making necessary further CLFV or LNV observations to establish full flavor breaking (a grid similar to Fig.~\ref{fig:LFV_grid} but with axes $L_\gamma$ and $L_\alpha-L_\beta$ can be drawn to make model-independent studies). It is straightforward but tedious and not particularly illuminating to extend the flowchart of Tab.~\ref{tab:flowchart} to LNV. (The processes in Tab.~\ref{tab:lnv} were also studied in connection to the Majorana neutrino mass matrix in Ref.~\cite{Hirsch:2006yk}.) Baryon number violation of course adds yet another dimension to the parameter space and can be explored completely analogously.

\section{Conclusion}

In summary, the observation of CLFV implies a breakdown of the approximate global flavor symmetry group $U(1)_{L_\mu-L_\tau}\times U(1)_{L_\mu+ L_\tau -2 L_e}$ in the charged-lepton sector. If \emph{one} process is observed, one linear combination of $L_\mu-L_\tau$ and $L_\mu+ L_\tau -2 L_e$ is broken, while the orthogonal one might still be conserved (up to tiny neutrino-mediated contributions). Depending on the process, this \emph{could} imply the existence of other testable CLFV channels (Tab.~\ref{tab:lfv}), but could also be an isolated process (Tab.~\ref{tab:lfv2}).
If \emph{two} (orthogonal) CLFV processes are observed, Fig.~\ref{fig:LFV_grid} can be used to predict additional processes, which might not necessarily be \emph{all} possible ones due to a possible remaining discrete subgroup.
CLFV is hence far more than a yes--no question, with up to \emph{three} qualitatively different discoveries required to establish that no flavor symmetry exists in the charged-lepton sector.
The discovery of lepton \emph{number} violation would practically imply lepton \emph{flavor} violation as well and thus has to be taken into account when interpreting data.
With a large number of experiments exploring untested parameter space, it is entirely possible that we see one or more discoveries within the next decade.

\section*{Acknowledgements}

I thank the organizers and participants of the \emph{XIIth Rencontres du Vietnam: Precise theory for precise experiments} for an inspiring conference that led to this work.
I am especially grateful to Sacha Davidson for comments on the manuscript and acknowledge support by the F.R.S.-FNRS as a postdoctoral researcher.

\bibliographystyle{utcaps_mod}
\bibliography{BIB}

\end{document}